\documentclass{sig-alternate-05-2015}
\PassOptionsToPackage{hyphens}{url}
\usepackage{hyperref}
\usepackage{subcaption}
\usepackage{lmodern}

\begin{document}

\title{Comparing Published Scientific Journal Articles\\to Their Pre-print Versions}

\numberofauthors{4}
\author{
%
\alignauthor
Martin Klein \\
	\affaddr{University of California\\Los Angeles}\\
	\affaddr{orcid.org/0000-0003-0130-2097}\\
	\email{martinklein@library.ucla.edu}
\alignauthor
Peter Broadwell \\
	\affaddr{University of California\\Los Angeles}\\
	\affaddr{orcid.org/0000-0003-4371-9472}\\
	\email{broadwell@library.ucla.edu}
\and
\alignauthor
Sharon E. Farb \\
	\affaddr{University of California\\Los Angeles}\\
	\affaddr{orcid.org/0000-0002-7655-1971}\\
	\email{farb@library.ucla.edu}
\alignauthor
Todd Grappone\\
	\affaddr{University of California\\Los Angeles}\\
	\affaddr{orcid.org/0000-0003-2218-7200}\\
	\email{grappone@library.ucla.edu}
}
\maketitle
\begin{abstract}
Academic publishers claim that they add value to scholarly communications by coordinating reviews and 
contributing and enhancing text during publication. These contributions come at a considerable cost: 
U.S. academic libraries paid $\$1.7$ billion for serial subscriptions in $2008$ alone.
Library budgets, in contrast, are flat and not able to keep pace with serial price 
inflation. We have investigated the publishers' value proposition by conducting a comparative study of 
pre-print papers and their final published counterparts. This comparison had two working assumptions: 
1) if the publishers' argument is valid, the text of a pre-print paper should vary measurably from 
its corresponding final published version, and 2) by applying standard similarity measures, we should 
be able to detect and quantify such differences. 
Our analysis revealed that the text contents of the scientific papers generally changed very little 
from their pre-print to final published versions. These findings contribute empirical indicators 
to discussions of the added value of commercial publishers and therefore should influence libraries' 
economic decisions regarding access to scholarly publications.  
\end{abstract}
%
%
%
%
%
%
%
%
%
%

\section{Introduction}
Academic publishers of all types claim that they add value to scholarly communications by coordinating 
reviews and contributing and enhancing text during publication. These contributions come at a 
considerable cost: U.S. academic libraries paid $\$1.7$ billion for serial subscriptions in $2008$ 
alone and this number continues to rise. Library budgets, in contrast, are flat and not able to keep 
pace with serial price inflation. Several institutions have therefore discontinued or significantly 
scaled back their subscription agreements with commercial publishers such as Elsevier and 
Wiley-Blackwell. At the University of California, Los Angeles (UCLA), we have investigated the 
publishers' value proposition by conducting a comparative study of pre-print papers and their final 
published counterparts. We have two working assumptions: 
\begin{enumerate}
\item If the publishers' argument is valid, the text of a pre-print paper should vary measurably from 
its corresponding final published version. 
\item By applying standard similarity measures, we should be able to detect and quantify such 
differences. 
\end{enumerate}
In this paper we present our preliminary results based on pre-print publications from arXiv.org and 
their final published counterparts obtained through subscriptions held by the UCLA Library. After 
matching papers via their digital object identifiers (DOIs), we applied comparative analytics and 
evaluated the textual similarities of components of the papers such as the title, abstract, and 
body. Our analysis revealed that the text contents of the papers in our test data set generally 
changed very little from their pre-print to final published versions; these results suggest that 
the vast majority of final published papers are largely indistinguishable from their pre-print 
versions. This work contributes empirical indicators to discussions of the value that academic 
publishers add to scholarly communication and therefore can influence libraries' economic decisions 
regarding access to scholarly publications.
\section{Global Trends in Scientific and Scholarly Publishing}
There are several global trends that are relevant and situate the focus of this research. The first 
is the steady rise in both cost and scope of the global science, technology and medicine (STM) 
publishing market. According to Michael Mabe and Mark Ware in their STM Report 
$2015$ \cite{ware:stmreport}, the global STM market in $2013$ was $\$25.2$ billion annually, 
with $40\%$ of this from journals ($\$10$ billion) and $68\%-75\%$ coming directly out of library 
budgets. Other relevant trends are the growing global research corpus \cite{bornmann:2015growth}, 
the steady rise in research funding \cite{ucla:accountability}, and the corresponding recent increase 
in open access publishing \cite{bjork:megajournals}. One longstanding yet infrequently mentioned 
factor is the critical contribution of faculty and researchers to the creation and establishment 
of journal content that is then licensed back to libraries to serve students, faculty 
and researchers. For example, a $2015$ Elsevier study (reported in \cite{ucla:accountability}) 
conducted for the University of California (UC) system showed that UC research publications 
accounted for $8.3\%$ of all research publications in the United States between $2009$ and 
$2013$ \textit{and the UC libraries purchased all of that research back from Elsevier}.  
\subsection{The Price of Knowledge}
While there are many facets to the costs of knowledge, the pricing of published scholarly 
literature is one primary component. Prices set by publishers are meant to maximize profit and 
therefore are determined not by actual costs, but by what the market will bear. According to the 
National Association of State Budget Officers, $24$ states in the U.S. had budgets in $2013$ with 
lower general fund expenditures in $FY13$ than just prior to the global recession in 
$2008$ \cite{budget:fy14}. Nearly half of the states therefore had not returned to pre-recession 
levels of revenue and spending.
\subsection{Rise in Open Access Publications }
Over the last several years there has been a significant increase in open access publishing and 
publications in STM. Some of this increase can be traced to recent U.S. federal guidelines and 
other funder policies that require open access publication. Examples include such policies at the 
National Institutes of Health, the Wellcome Trust, and the Howard Hughes Medical Center. 
Bo-Christer Bj{\"o}rk et al. \cite{bjork:openaccess} found that in $2009$, approximately $25\%$ 
of science papers were open access. By $2015$, another study by Hammid R. Jamali and Maijid 
Nabavi \cite{jamali:openaccess} found that $61.1\%$ of journal articles were freely available 
online via open access. 
\subsection{Pre-print versus Final Published Versions and the Role of Publishers}
In this study, we compared paper pre-prints from arXiv.org to the corresponding final published 
versions of the papers. For comparison, the annual budget for arXiv.org is set at $\$826,000$ 
for $2013-2017$. While it is not possible to determine the precise corresponding costs for 
commercial publishing, the National Center for Education Statistics $2013$ found that the market 
for English language STM journals was approximately $\$10$ billion dollars annually. It therefore 
seems safe to say that the costs for commercial publishing are orders of magnitude larger than the 
costs for an organization such as arXiv.org.

Michael Mabe \cite{mabe:ermh} describes the publishers' various roles as including, but not limited 
to entrepreneurship, copyediting, tagging, marketing, distribution, and e-hosting. The focus of 
the study presented here is on the publishers' contributions to the content of the materials they 
publish (specifically copyediting and other enhancements to the text) and how and to 
what extent, if at all, the content changes from the pre-print to the final published version of 
a publication.
\section{Data Gathering}
Comparing pre-prints to final published versions of a significant corpus of scholarly articles 
required obtaining the contents of both versions of each article in a format that could be analyzed 
as full text and parsed into component sections (title, abstract, body) for more detailed comparisons. 
The most accessible source of such materials proved to be \texttt{arXiv.org}, an open-access digital 
repository owned and operated by Cornell University and supported by a consortium of institutions. 
At the time of writing, arXiv.org hosts over $1.1$ million academic pre-prints, most written in fields 
of physics and mathematics and uploaded by their authors to the site within the past $20$ years. The 
scope of arXiv.org also enabled us to identify and obtain a sufficiently large comparison corpus of 
corresponding final published versions in scholarly journals to which our institution has access via 
subscription.
\subsection{arXiv.org Corpus}
Gathering pre-print texts from arXiv.org proceeded via established public interfaces for machine 
access to the site data, respecting their discouragement of indiscriminate automated 
downloads.\footnote{\url{https://arxiv.org/help/robots}}

We first downloaded metadata records for all articles available from arXiv.org through February 
of $2015$ via the site's Open Archives Initiatives Protocol for Metadata Harvesting (OAI-PMH) 
interface.\footnote{\url{http://export.arxiv.org/oai2?verb=Identify}} We received $1,015,440$ 
records in all, which provided standard Dublin Core metadata for each article, including its title 
and authors, as well as other useful data for subsequent analysis, such as the paper's disciplinary 
category within arXiv.org and the upload dates of its versions (if the authors submitted more than 
one version). The metadata also contained the text of the abstract for most articles. Because the 
abstracts as well as the article titles often contained text formatting markup, however, we 
preferred to use instances of these texts that we derived from other sources, such as the PDF version 
of the paper, for comparison purposes (see below).

arXiv.org's OAI-PMH metadata record for each article contains a field for a DOI, which we used as 
the key to match pre-print versions of articles to their final published versions. arXiv.org does 
not require DOIs for submitted papers, but authors may provide them voluntarily. $452,017$ article 
records in our initial metadata set ($44.5\%$) contained a DOI. Working under the assumption that 
the DOIs are correct and sufficient to identify the final published version of each article, we 
then queried the publisher-supported CrossRef citation linking service 
service\footnote{\url{https://github.com/CrossRef/rest-api-doc/blob/master/rest_api.md}} to 
determine whether the full text of the corresponding published article would be available for 
download via UCLA's institutional journal subscriptions.

To begin accumulating full articles for text comparison, we downloaded PDFs of every pre-print 
article from arXiv.org with a DOI that could be matched to a full-text published version
accessible through subscriptions held by the UCLA Library.
Our initial query indicated that up to $12,666$ final published versions would be accessible in 
this manner. 
The main reason why this number is fairly low is that, at the time of writing, the above mentioned 
CrossRef API is still in its early stages and only few publishers have agreed to making their 
articles available for text and data mining via the API. 
However, while this represented a very small proportion of all papers with DOI-associated pre-prints 
stored in arXiv.org, the resulting collection nevertheless proved sufficient for a detailed 
comparative analysis.

The downloads of pre-prints took place via arXiv.org's bulk data access service, which facilitates 
the transfer of large numbers of articles as PDFs or as text markup source files and images, 
packaged into .tar archives, from an Amazon S3 account. Bandwidth fees are paid by the requesting 
party.\footnote{\url{https://arxiv.org/help/bulk_data_s3}} This approach only yields the most 
recent uploaded version of each pre-print article, however, so for analyses involving earlier 
uploaded versions of pre-print articles, we relied upon targeted downloads of earlier article 
versions via arXiv.org's public web interface.
\subsection{Corpus of Matched Articles}
Obtaining the final published versions of articles involved querying the CrossRef API to find a
full-text download URL for a given DOI. Most of the downloaded files ($96$\%) arrived in one of 
a few standard XML markup formats; the rest were in PDF format. Due to missing or incomplete target 
files, $464$ of the downloads failed entirely, leaving us with $12,202$ published versions for 
comparison. The markup of the XML files contained, in addition to the full text, metadata entries 
from the publisher. Examination of this data revealed that the vast majority ($99\%$) of articles 
were published between $2003$ and $2015$. This time range intuitively makes sense as DOIs did not 
find widespread adoption with commercial publishers until the early $2000s$. The data also shows 
that most of the obtained published versions ($96$\%) were published by Elsevier.
\begin{figure}[h!]
\center
\includegraphics[scale=0.17]{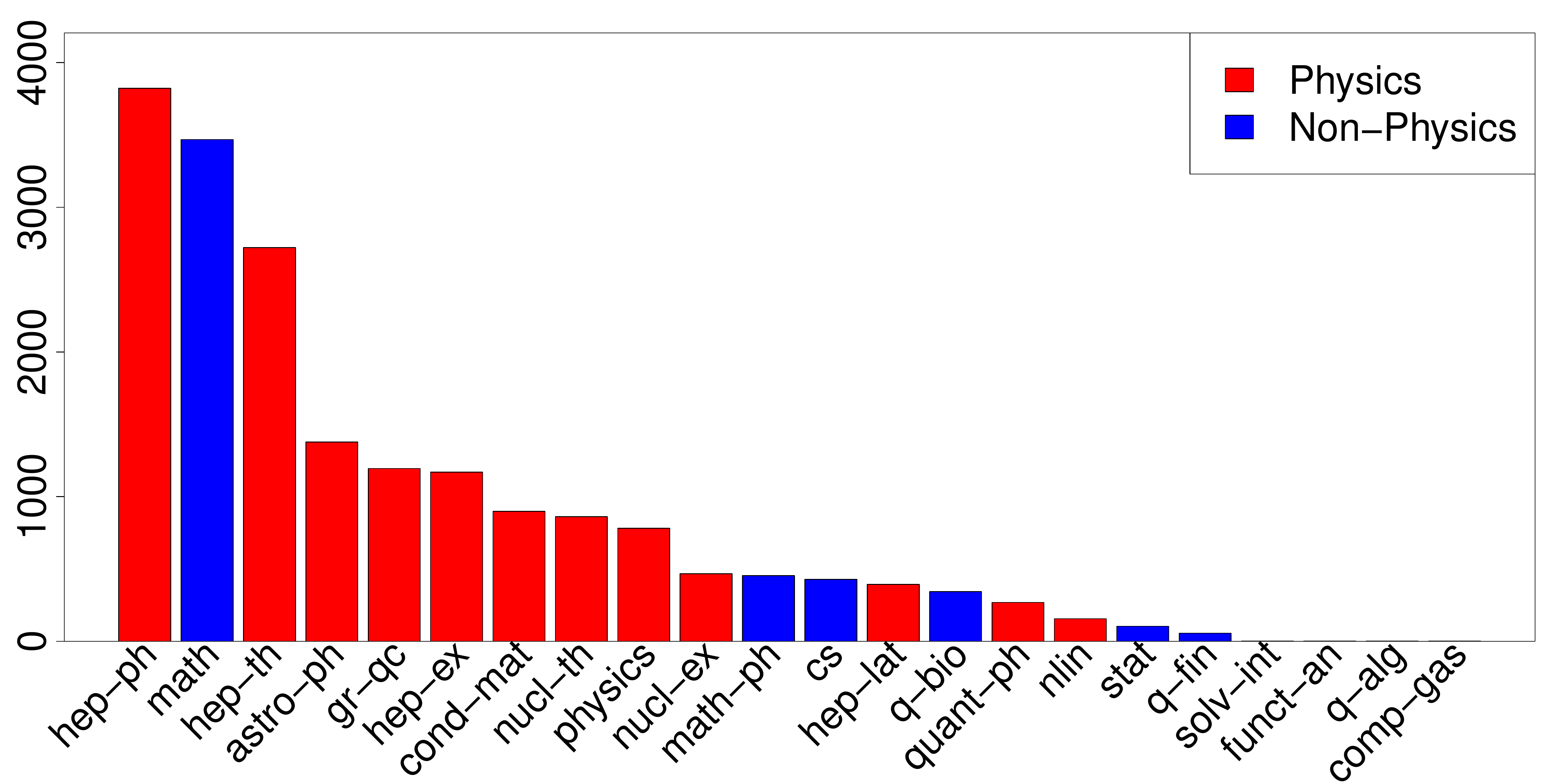}
\caption{arXiv.org categories of matched articles}
\label{fig:arxiv_categories}
\end{figure}

The disciplines of articles in arXiv.org are dominated by physics, mathematics, statistics, and 
computer science. It is therefore not surprising to find a very similar distribution of categories 
in our corpus of matched articles as shown in Figure \ref{fig:arxiv_categories}.
\subsection{Data Preparation}
For this study, we compared the texts of the titles, abstracts, and body sections of the pre-print 
and final published version of each paper in our data set. Being able to generate these sections 
for most downloaded papers therefore was a precondition of this analysis.

All of the pre-print versions and a small minority of final published papers were downloaded in 
PDF format. To identify and extract the sections of these papers, we used the 
GROBID\footnote{\url{https://github.com/kermitt2/grobid}} library, which employs trained conditional 
random field machine learning algorithms to segment structured scholarly texts, including article 
PDFs, into XML-encoded text.

The markup tags of the final published papers downloaded in XML format usually identified quite 
plainly their primary sections. A small proportion of such papers, however, did not contain a 
demarcated body section in the XML and instead only provided the full text of the papers. Although 
it is possible to segment these texts further via automatic scholarly information extraction tools 
such as ParsCit,\footnote{\url{http://aye.comp.nus.edu.sg/parsCit/}} which use trained conditional 
random field models to detect sections probabilistically, for the present study we elected simply 
to omit the body sections of this small number of papers from the comparison analysis.

As noted above, the GROBID software used to segment the PDF papers was probabilistic in its approach, 
and although it was generally quite effective, it was not able to isolate all sections (title, 
abstract, body) for approximately $10-20$\% of the papers in our data set. This situation, combined 
with the aforementioned irregularities in the XML of a similar proportion of final published papers, 
meant that the number of corresponding texts for comparison varied considerably by section. Thus, 
for our primary comparison of the latest pre-print version uploaded to arXiv.org to its final 
published version, we were able to compare directly $10,900$ titles and abstract sections, and 
$9,399$ body sections.

The large variations in formatting of the references sections (also called the ``tail'') as extracted 
from the raw downloaded XML and the parsed PDFs, however, precluded a systematic comparison of that 
section. We leave such an analysis for future work. A further consequence of our text-only analysis 
was that the contents of images were ignored entirely, although figure captions and the text contents 
of tables usually could be compared effectively.
\section{Analytical Methods} \label{subsec:txt_comp_meth}
We applied several text comparison algorithms to the corresponding sections of the pre-print and 
final published versions of papers in our test data set. These algorithms, described in detail below, 
were selected to quantify different notions of ``similarity'' between texts. When possible, we 
normalized the output values of each algorithm to lie between $1$ and $0$, with $1$ indicating that 
the texts were effectively identical, and $0$ indicating complete dissimilarity. Different algorithms 
necessarily measured any apparent degree of dissimilarity in different ways, so the outputs of the 
algorithms cannot be compared directly, but it is nonetheless valid to interpret the aggregation 
of these results as a general indication of the overall degree of similarity between two texts 
along several different axes of comparison.
\subsection{Editorial Changes}
The well-known Levenshtein edit distance metric \cite{levenshtein:edit_distance}
calculates the number of character insertions, deletions, and substitutions necessary to convert one 
text into another. It thus provides a useful quantification of the amount of editorial intervention 
--- performed either by the authors or the journal editors --- that occurs between the pre-print and 
final published version of a paper. Our work used the edit ratio calculation as provided in the 
Levenshtein Python C Implementation 
Module,\footnote{\url{https://pypi.python.org/pypi/python-Levenshtein/0.11.2}} which subtracts the 
edit distance between the two documents from their combined length in characters and divides this 
amount by their aggregate length, thereby producing a value between $1$ (completely similar) and 
$0$ (completely dissimilar).
\subsection{Length Similarity}
The degree to which the final published version of a paper is shorter or longer than the pre-print 
constitutes a much less involved but nonetheless revealing comparison metric. To calculate this 
value, we divided the absolute difference in length between both papers by the length of the longer 
paper and subtracted this value from $1$. Therefore, two papers of the same length will receive a 
similarity score of $1$; this similarity score is $0.5$ if one paper is twice as long as the other, 
and so on. It is also possible to incorporate the polarity of this change by adding the length 
ratio to $0$ if the final version is longer, and subtracting it from $0$ if the pre-print is longer.
\begin{figure*}[t!]
\center
\includegraphics[scale=0.5]{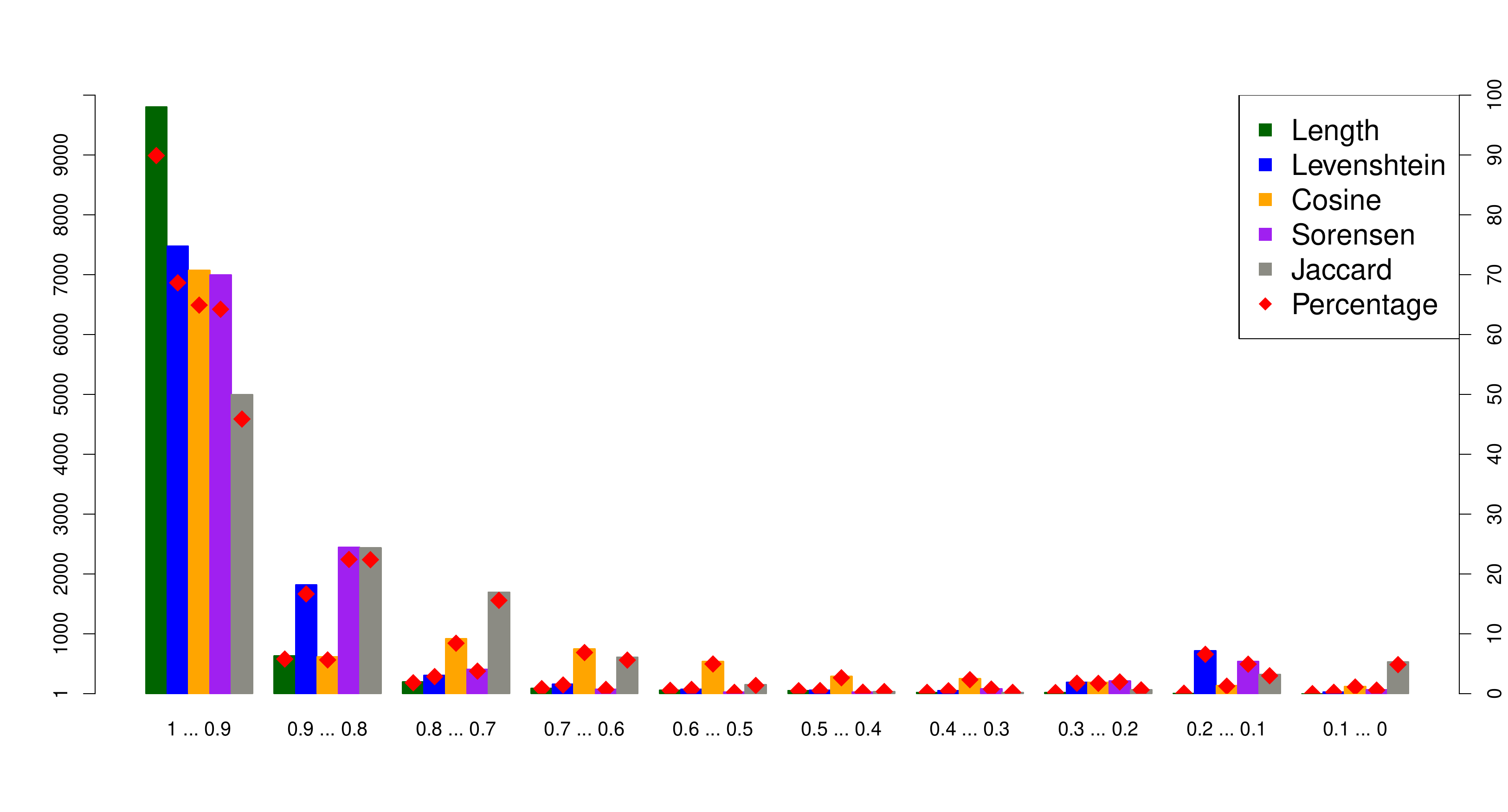}
\caption{Comparison results for titles}
\label{fig:title_histo}
\end{figure*}
\subsection{String Similarity}
Two other fairly straightforward, low-level metrics of string similarity that we applied to the paper 
comparisons were the Jaccard and S\o rensen indices, which consider only the sets of unique characters 
that appear in each text. The S\o rensen similarity \cite{sorensen:index} was calculated by doubling 
the number of unique characters shared between both texts (the intersection) and dividing this by 
the combined sizes of both texts' unique character sets. 

The Jaccard similarity calculation \cite{jaccard:index} is the size of the intersection (see above) 
divided by the total number of unique characters appearing in either the pre-print or final published 
version (the union). 

Implementations of both algorithms were provided by the standard Python string distance 
package.\footnote{\url{https://pypi.python.org/pypi/Distance/}}
\subsection{Semantic Similarity}
Comparing overall lengths, shared character sets, and even edit distances between texts does not 
necessarily indicate the degree to which the meaning of the texts --- that is, their semantic 
content --- actually has changed from one version to another. To estimate this admittedly more 
subjective notion of similarity, we calculated the pairwise cosine similarity between the pre-print 
and final published texts. Cosine similarity can be described intuitively as a measurement of how 
often significant words occur in similar quantities in both texts, normalized by the lengths of 
both documents \cite{pang:introdm}. The actual procedure used for this study involved removing 
common English ``stopwords'' from each document, then applying the Porter stemming 
algorithm \cite{porter:algo} to remove suffixes and thereby merge closely related words, before 
finally applying the pairwise cosine similarity algorithm implemented in the Python scikit-learn 
machine learning package\footnote{\url{http://scikit-learn.org/stable/}} to the resulting term 
frequency lists. Because this implementation calculates only the similarity between two documents 
considered in isolation, instead of within the context of a larger corpus, it uses raw term counts, 
rather than term-frequency/inverse document frequency (TF-IDF) weights.
\section{Experiment Results}
We calculated the similarity metrics described above for each pair of corresponding pre-print and 
final published papers in our data set, comparing titles, abstracts, and body sections when available. 
From the results of these calculations, we generated visualizations of the similarity distributions 
for each metric. Subsequent examinations and analyses of these distributions provided novel insights 
into the question of how and to what degree the text contents of scientific papers may change from 
their pre-print instantiations to the final published version. Because each section of a publication 
differs in its purpose and characteristics (e.g., length, standard formatting) and each metric 
addresses the notion of similarity from a different perspective, we present the results of our 
comparisons per section (title, abstract, and body), subdivided by comparison metric.
\begin{figure*}[ht!]
\center
\includegraphics[scale=0.5]{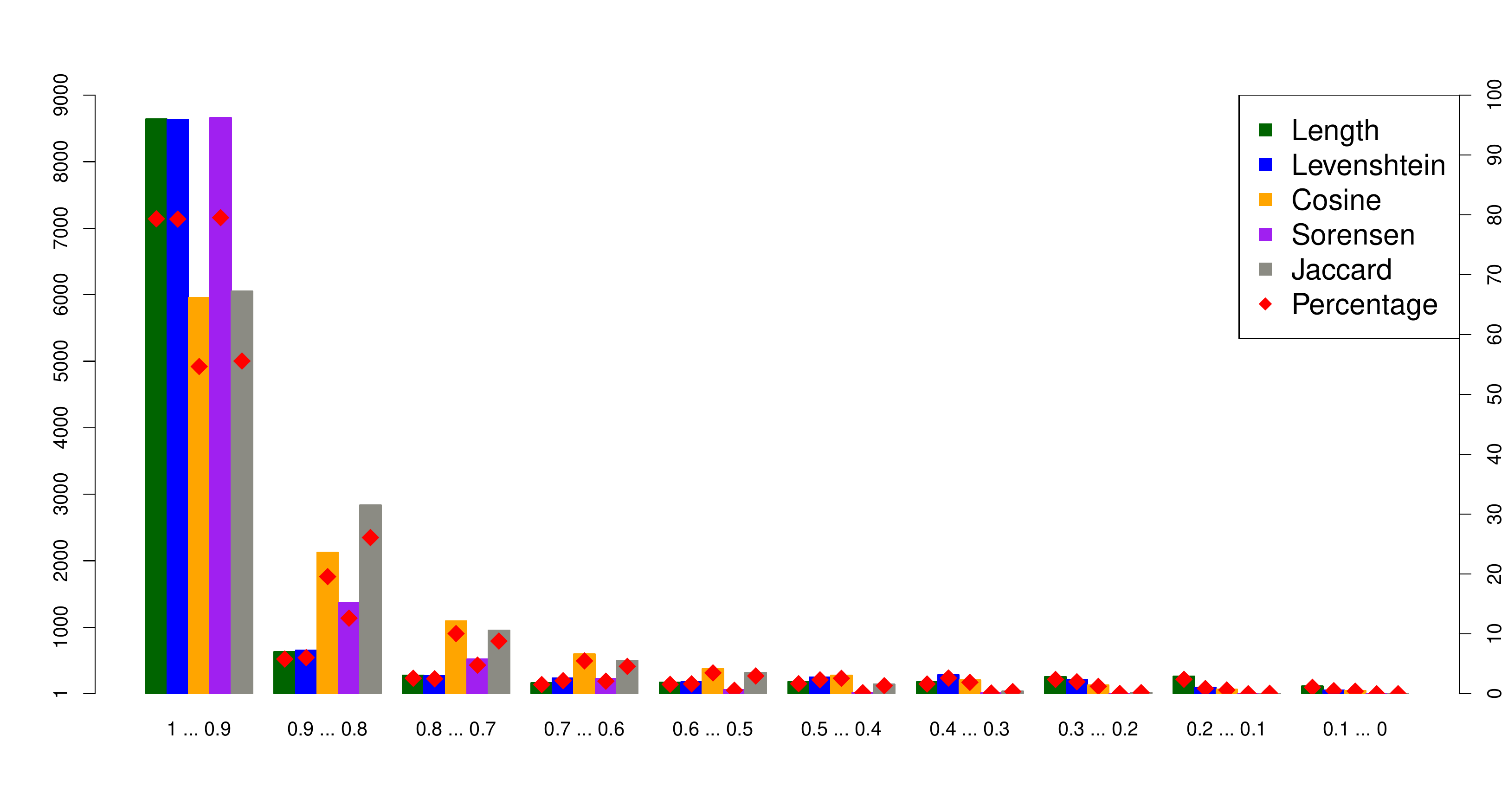}
\caption{Comparison results for abstracts}
\label{fig:abstract_histo}
\end{figure*}
\subsection{Title Analysis}
First, we analyzed the papers' titles. A title is usually much shorter (fewer characters) than a 
paper's abstract and its body. That means that even small changes to the title would have a large 
impact on the similarity scores based on length ratio and Levenshtein distance. Titles also often 
contain salient keywords describing the overall topic of the paper. If those keywords were changed, 
removed or new ones added, the cosine similarity value would drop.

Figure \ref{fig:title_histo} shows the comparison of results of all five text similarity measures 
applied to titles. Since all measures are normalized, they return values between $0$ and $1$. Values 
closer to $0$ represent a high degree of dissimilarity and values close to $1$ indicate a very high 
level of similarity of the analyzed text. The horizontal x-axis in Figure \ref{fig:title_histo} 
shows results aggregated into ten bins. The bin with the largest values between $0.9$ and $1.0$ is 
located on the far left of the axis followed by the bin with values between $0.9$ and $0.8$ and so 
on. The bin with values between $0$ and $0.1$ can be found on the right end of the x-axis. Each 
bin contains five columns, each of which represents one applied similarity measure. A column's 
height indicates the number of articles whose title similarity scores fall into the corresponding 
bin. The height of a column refers to the left y-axis and is shown in absolute numbers. The red 
diamond-shaped point in each column indicates the relative proportion of articles in the entire 
corpus that is represented by the corresponding column. The value of a diamond refers to the 
right y-axis where the percentage is shown.

Figure \ref{fig:title_histo} shows a dominance of the top bin. The vast majority of titles have a 
very high score in all applied similarity measures. Most noticeably, almost $10,000$ titles 
(around $90\%$ of all titles) are of very similar length, with a ratio value between $0.9$ and $1$. 
The remaining $10\%$ fall into the next bin with values between $0.8$ and $0.9$.
A very similar observation can be made for the Levenshtein distance and the S\o rensen value. About 
$70\%$ of those values fall into the top bin and the majority of the remaining values (around $20\%$)
land between $0.8$ and $0.9$.  
The cosine similarity is also dominated by values in the top bin (around $70\%$) but the remaining 
values are more distributed across the second, third, fourth, and fifth bin.
Just about half of all Jaccard values can be seen in the top bin and most of the remainder is split 
between the second ($25\%$) and the third bin ($20\%$). In many cases, this metric is registering 
low-level but systematic differences in character use between the pre-print and final published 
versions as filtered through the download methods described above: for example, a pre-print may 
consistently use em-dashes (--), whereas the published version uses only hyphens (-). This sensitivity 
of the Jaccard similarity score to subtle changes in the unique character sets in each text is 
apparent for other sections as well.

The results of this comparison, in particular the fact that the majority of values fall between 
$0.9$ and $1$, provide very strong indicators that titles of scholarly articles do not change 
noticeably between the pre-print and the final published version. Even though 
Figure \ref{fig:title_histo} shows a small percentage of titles exhibiting a rather low level of 
similarity, with Levenshtein and S\o rensen values between $0.1$ and $0.2$, the overall similarity 
of titles is very high.
\begin{figure*}[ht!]
\center
\includegraphics[scale=0.5]{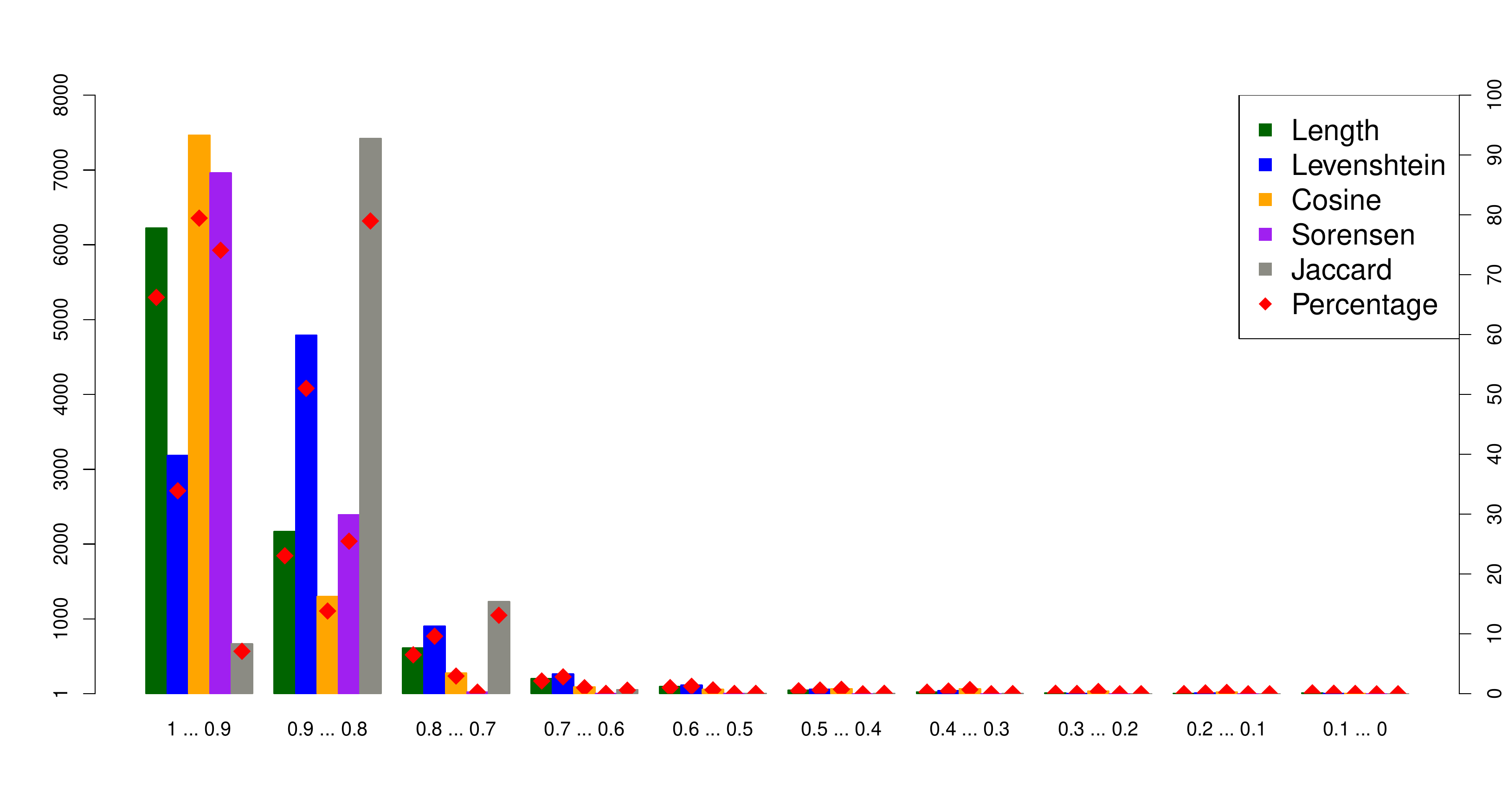}
\caption{Comparison results for body sections}
\label{fig:body_histo}
\end{figure*}
%
%
\newpage
\subsection{Abstract Analysis}
The next section we compared was the papers' abstracts. An abstract can be seen as a very short 
version of the paper. It often gives a brief summary of the problem statement, the methods applied, 
and the achievements of the paper. As such, an abstract usually is longer than the paper's title 
(in number of characters) and provides more context. Intuitively, it seems probable that we would 
find more editorial changes in longer sections of the pre-print version of an article compared to 
its final published version.  However, a potentially increased number of editorial changes alone 
does not necessarily prove dissimilarity between longer texts. We expect similarity measures based 
on semantic features such as cosine similarity to be more reliable here.

Figure \ref{fig:abstract_histo} shows the comparative results for all abstracts. The formatting of 
the graph is the same as previously described for Figure \ref{fig:title_histo}. To our surprise, 
the figure is dominated by the high frequency of values between $0.9$ and $1$ across all similarity 
measures. More than $8,500$ abstracts (about $80\%$) have such scores for their length ratio, 
Levenshtein distance, and S\o rensen index. $6\%$ of the remaining length ratio and Levenshtein 
distance values as well as $13\%$ of the remaining S\o rensen index values fall between $0.8$ and 
$0.9$. The remaining pairs are distributed across all other bins.
The cosine similarity and Jaccard index values are slightly more distributed. About $5,000$ abstracts 
($55\%$) fall into the top bin, $20\%$ and $26\%$ into the second, and $10\%$ and $9\%$ into the 
third bin, respectively.

Not unlike our observations for titles, the algorithms applied to abstracts predominantly return 
values that indicate a very high degree of similarity. Figure \ref{fig:abstract_histo} shows that 
more than $90\%$ of abstracts score $0.6$ or higher, regardless of the text similarity measure 
applied. It is also worth pointing out that there is no noticeable increased frequency of values 
between $0.1$ and $0.2$ as previously seen when comparing titles (Figure \ref{fig:title_histo}). 
\subsection{Body Analysis} \label{sec:body_analysis}
The next section we extracted from our corpora of scholarly articles and subjected to the text 
similarity measures is the body of the text. This excludes the title, the author(s), the abstract, 
and the reference section. This section is, in terms of number of characters, the longest of our 
three analyzed sections. We therefore consider scores resulting from algorithms based on editorial 
changes to be less informative for this comparison. In particular, a finding such as ``The body 
of article $A_2$ contains $10\%$ fewer characters than the body of article $A_1$'' would not provide 
any reliable indicators of the similarity between the two articles $A_1$ and $A_2$. Algorithms based 
on semantic features, such as the cosine similarity, on the other hand, provide stronger indicators 
of the similarity of the compared long texts. More specifically, cosine values are expected to be 
rather low for very dissimilar bodies of articles.

The results of this third comparison can be seen in Figure \ref{fig:body_histo}. The height of the 
bar representing the cosine similarity is remarkable. Almost $7,500$ body sections of our compared 
scholarly articles, which is equivalent to $80\%$ of the entire corpus, have a cosine score that 
falls in the top bin with values between $0.9$ and $1$. $14\%$ have a cosine value that falls into 
the second and $3\%$ fall into the third bin. Values of the S\o rensen index show a very similar 
pattern with $74\%$ in the top bin and $25\%$ in the second. In contrast, only $7\%$ of articles' 
bodies have Jaccard index values falling into the top bin. The vast majority of these scores, 
$79\%$, are between $0.8$ and $0.9$ and another $13\%$ are between $0.7$ and $0.8$.
It is surprising to see that even the algorithms based on editorial changes provide scores mostly 
in the top bins. Of the length ratio scores, $66\%$ fall in the top bin and $23\%$ in the second 
bin. The Levenshtein distance shows the opposite proportions: $34\%$ are in the top and $51\%$ 
belong to the second bin.

The dominance of bars on the left hand side of Figure \ref{fig:body_histo} provides yet more evidence
that pre-print articles of our corpus and their final published version do not exhibit many features 
that could distinguish them from each other, neither on the editorial nor on the semantic level. 
$95\%$ of all analyzed body sections have a similarity score of $0.7$ or higher in any of the 
applied similarity measures. 
\subsection{Publication Dates} \label{subsec:pub_dates}
\begin{figure}[t!]
\center
\includegraphics[scale=0.17]{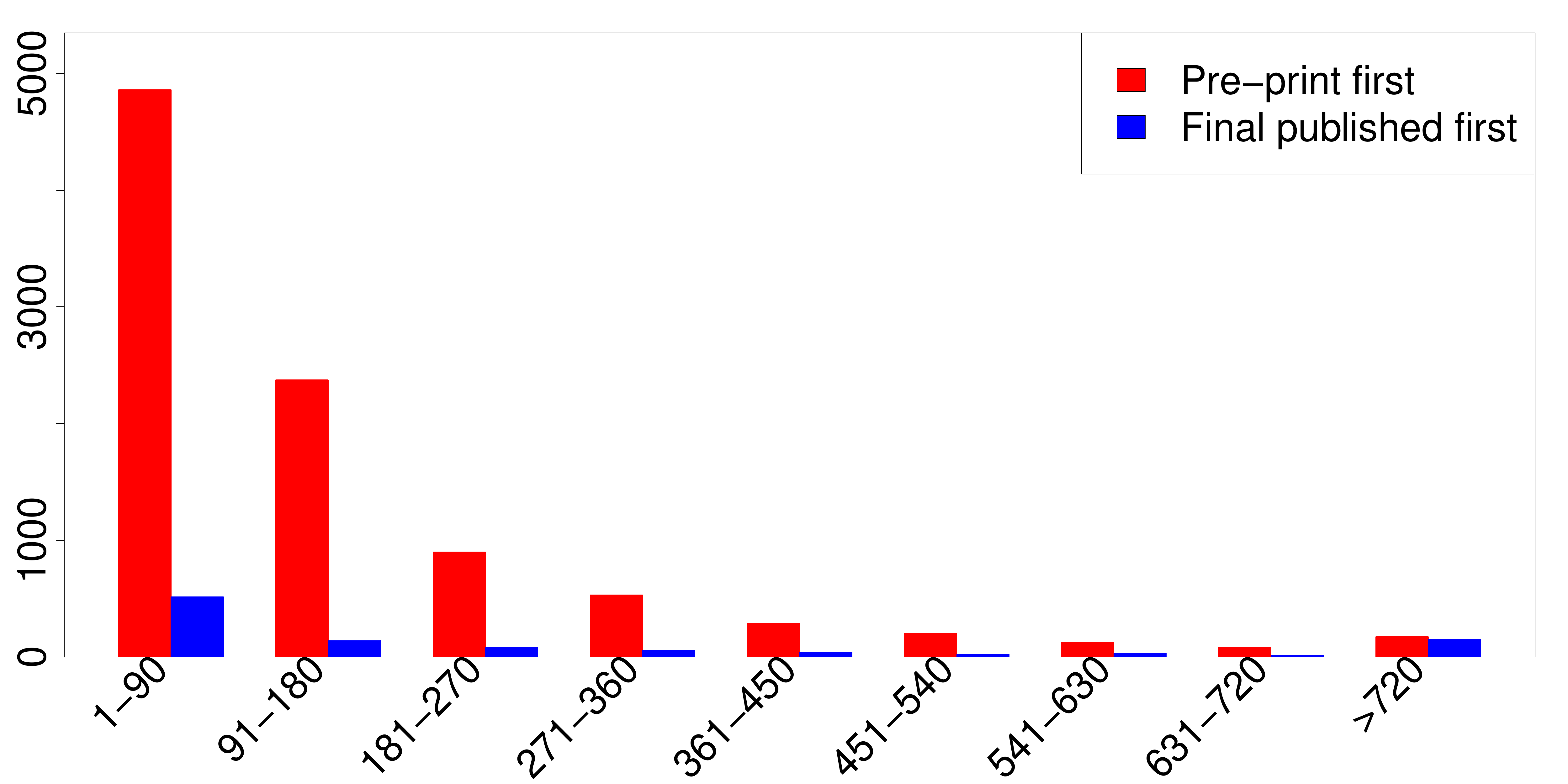}
\caption{Numbers of articles first appearing in the specified venue, given the date of the last pre-print upload and the commercial publication date, binned by the number of days between them}
\label{fig:pub_dates}
\end{figure}

The above results provide strong indicators that there is hardly any noticeable difference between 
the pre-print version of a paper and its final published version. However, the results do not show
which version came first. In other words, consider the two possible scenarios:
\begin{enumerate}
\item Papers, after having gone through a rigorous peer review process, are published by a commercial 
publisher first and then, as a later step, uploaded to arXiv.org. In this case the results of our 
text comparisons described above would not be surprising, as the pre-print versions would merely be 
a mirror of the final published ones. There would be no apparent reason to deny publishers all credit 
for peer review, copyediting, and the resulting publication quality of the articles.
\item Papers are uploaded to arXiv.org first and later published by a commercial publisher. If this 
scenario is dominant, our comparison results would suggest that any changes in the text due to 
publisher-initiated copyediting are hardly noticeable.
\end{enumerate}
\begin{figure*}
\center
\begin{subfigure}[a]{1\textwidth}
	\center
	\includegraphics[scale=0.4]{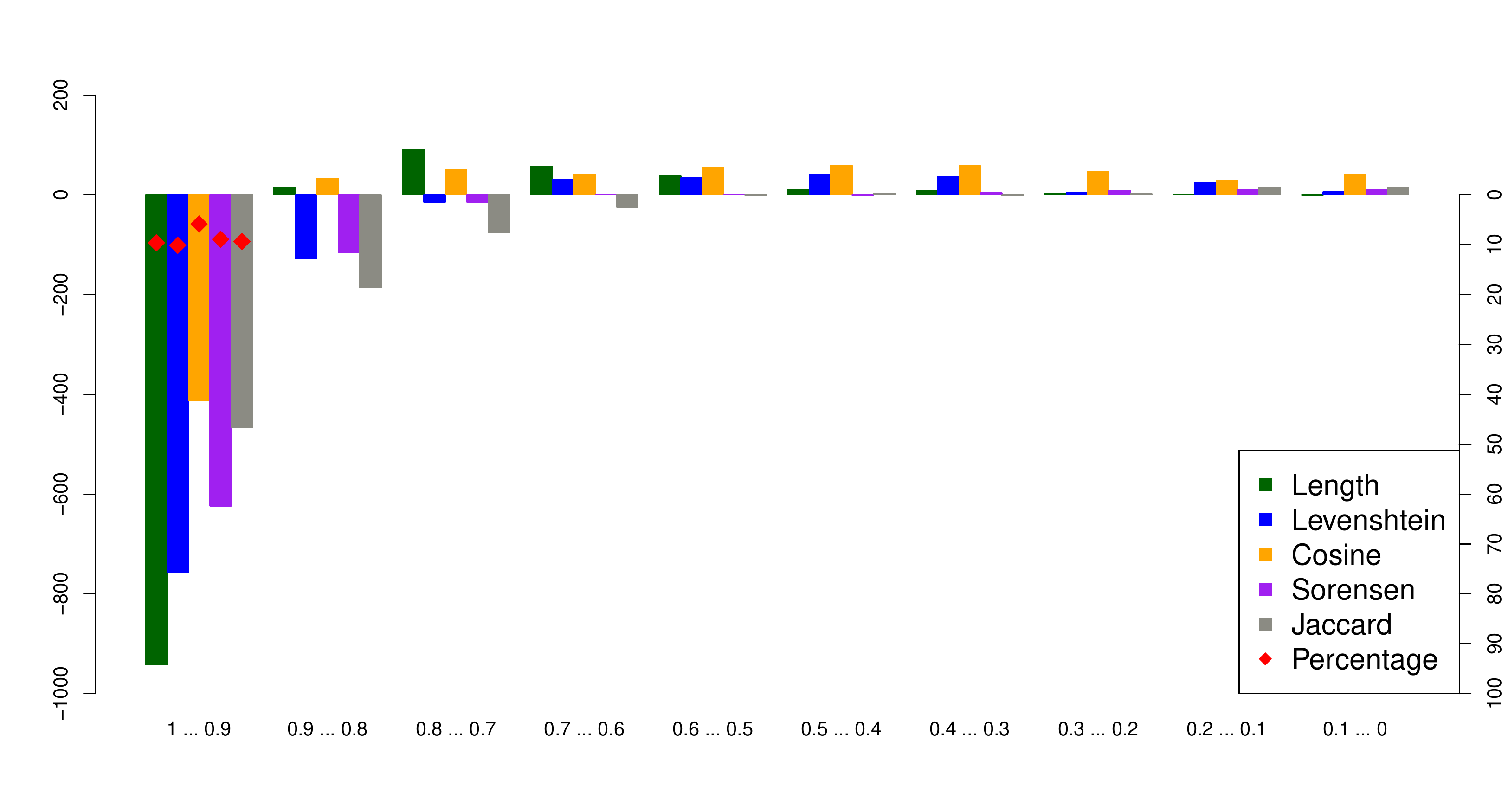}
	\caption{Title deltas}
	\label{subfig:title_delta}
\end{subfigure}
\begin{subfigure}[a]{1\textwidth}
	\center
	\includegraphics[scale=0.4]{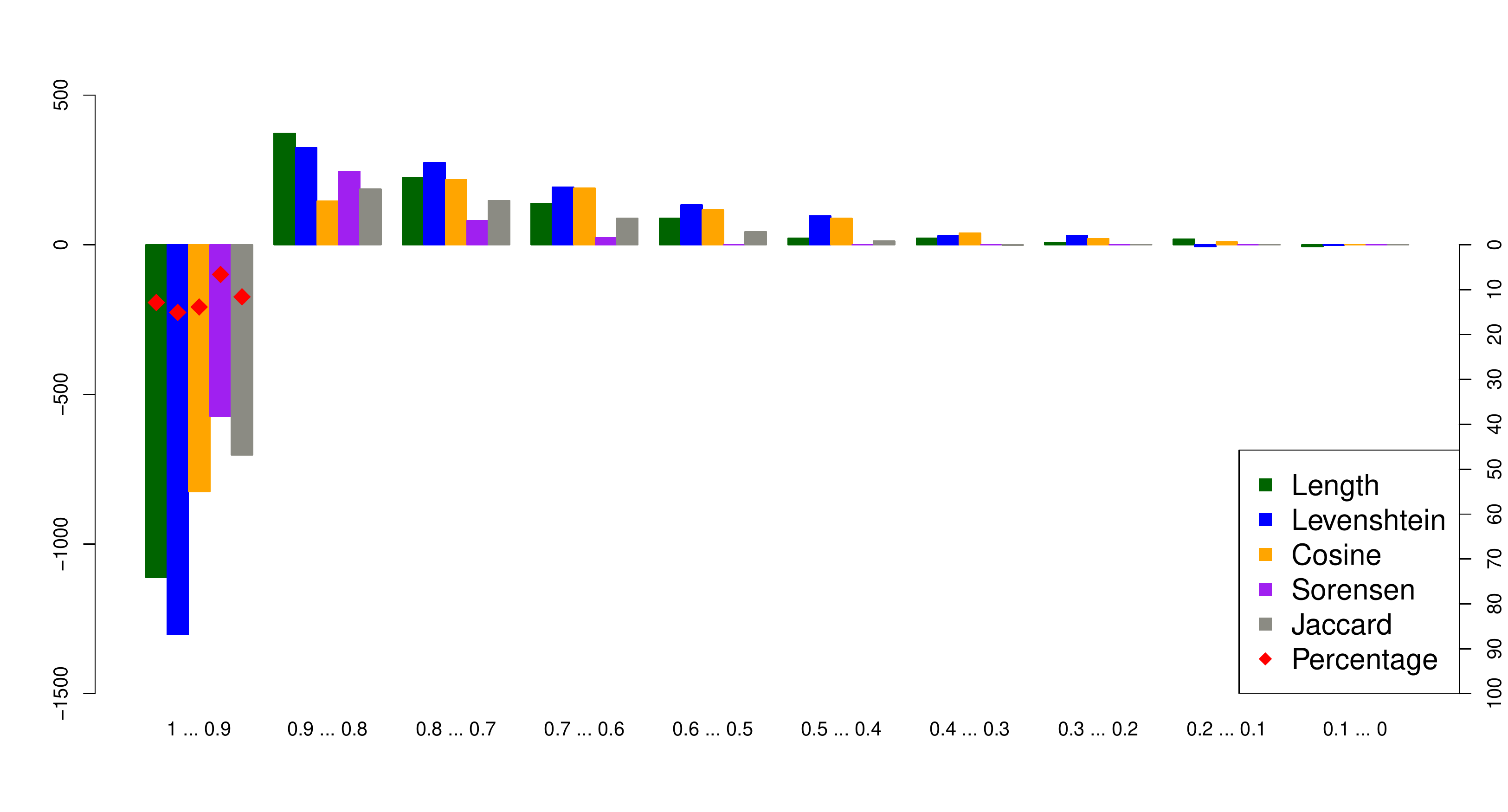}
	\caption{Abstract deltas}
	\label{subfig:abstract_delta}
\end{subfigure}
\begin{subfigure}[a]{1\textwidth}
	\center
	\includegraphics[scale=0.4]{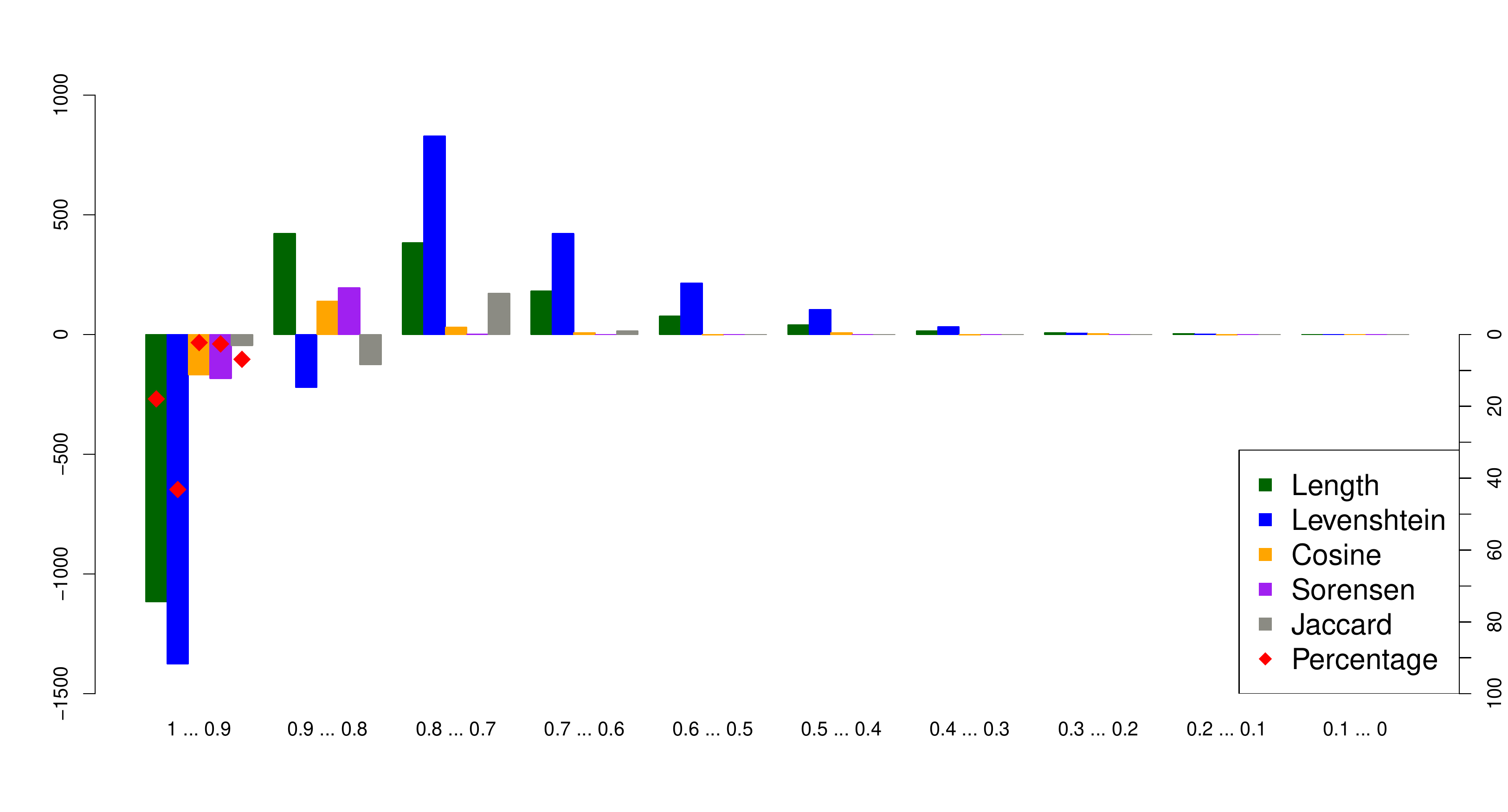}
	\caption{Body deltas}
	\label{subfig:body_delta}
\end{subfigure}
\caption{Deltas of paper section comparison bins for the five similarity metrics in the legend, 
showing the differences in the bin values relative to those in Figures 
\ref{fig:title_histo}, \ref{fig:abstract_histo}, and \ref{fig:body_histo}}
\label{fig:deltas}
\end{figure*}
Figure \ref{fig:pub_dates} shows the order of appearance in arXiv.org versus commercial venues for 
all articles in our corpus, comparing the publication date of each article's final published version 
to the date of its latest upload to arXiv.org. Red bars indicate the amount of articles (absolute 
values on the y-axis) that were first upload to arXiv.org, and blue bars stand for articles published 
by a commercial publisher before they appeared in arXiv.org. Each pair of bars is binned into a time 
range, shown on the x-axis, that indicates approximately how many days passed between the article's 
appearance in the indicated first venue and its appearance in the second venue.
Figure \ref{fig:pub_dates} show clear evidence that the vast majority of our articles ($90\%$) were 
published in arXiv.org first. Therefore our argument for the second scenario holds. We can only 
speculate about the causes of certain time windows' prominence within the distribution, but it may 
be related to turn-around times of publishers between submission and eventual publication.
\begin{table*}[ht!]
\center
\caption{Deltas of the proportions of paper titles in the entire corpus that belong to the 
specified bins for each comparison metric, giving the differences between the proportional 
values (red diamonds) in Figure \ref{fig:title_histo} and those generated when each article's 
first uploaded version in arXiv.org is considered (shown in Figure \ref{subfig:title_delta})}
\begin{tabular}{|c||c|c|c|c|c|c|c|c|c|c|} \hline
Measure & Bin $1$ & Bin $2$ & Bin $3$ & Bin $4$ & Bin $5$ & Bin $6$ & Bin $7$ & Bin $8$ & Bin $9$ & Bin $10$ \\ \hline \hline
len&  -9.61&  2.37& 45.73& 61.29& 61.29& 20.00& 32.00&  9.52& 10.00& -50.00 \\ \hline
lev& -10.12& -7.04& -4.81& 19.50& 43.04& 72.41& 68.52&  2.60&  3.48&  23.08 \\ \hline
cos&  -5.84&  5.36&  5.43&  5.47& 10.15& 20.27& 22.75& 24.87& 21.01&  33.33 \\ \hline
sor&  -8.91& -4.70& -3.64&  1.25&  0.00& -3.45&  4.71&  4.17&  2.04&  14.93 \\ \hline
jac&  -9.34& -7.62& -4.47& -4.08& -0.66&  7.50& -8.00&  2.82&  4.88&   3.02 \\ \hline
\end{tabular}
\label{tab:title_delta}
\end{table*}
\begin{table*}
\center
\caption{Proportional deltas for abstracts (shown in Figure \ref{subfig:abstract_delta})}
\begin{tabular}{|c||c|c|c|c|c|c|c|c|c|c|} \hline
Measure & Bin $1$ & Bin $2$ & Bin $3$ & Bin $4$ & Bin $5$ & Bin $6$ & Bin $7$ & Bin $8$ & Bin $9$ & Bin $10$ \\ \hline \hline
len& -12.87& 58.95&  79.36& 83.23& 51.46& 12.09& 12.36&  3.08&  7.25& -5.13 \\ \hline
lev& -15.08& 49.39& 100.73& 81.43& 74.30& 38.34& 10.49& 14.68& -6.25& -3.64 \\ \hline
cos& -13.86&  6.85&  19.91& 31.66& 30.69& 31.54& 18.84& 15.67& 13.04& -2.17 \\ \hline
sor&  -6.63& 17.89&  15.55& 10.53& -1.56& -5.26&  0&  0&  0&  0  \\ \hline
jac& -11.60&  6.55&  15.45& 17.84& 13.62&  8.22& -5.13&  0&  0&  0 \\ \hline
\end{tabular}
\label{tab:abstract_delta}
\end{table*}
\begin{table*}
\center
\caption{Proportional deltas for body sections (shown in Figure \ref{subfig:body_delta})}
\begin{tabular}{|c||c|c|c|c|c|c|c|c|c|c|} \hline
Measure & Bin $1$ & Bin $2$ & Bin $3$ & Bin $4$ & Bin $5$ & Bin $6$ & Bin $7$ & Bin $8$ & Bin $9$ & Bin $10$ \\ \hline \hline
len& -17.93& 19.43& 62.58&  90.05&  79.38&  85.11& 60.00& 88.89& 133.33&   0 \\ \hline
lev& -43.14& -4.61& 91.59& 158.65& 183.76& 177.97& 74.42& 83.33&  14.29&   0 \\ \hline
cos&  -2.25& 10.62& 10.79&   7.37&  -1.64&  10.00& -3.03& 11.11&  -4.76&   0 \\ \hline
sor&  -2.63&  8.14&  3.85&   0&   0&    0&  0&   0&    0&   0 \\ \hline
jac&  -6.90& -1.70& 13.87&  25.00&   0&   0&  0&  0&    0&   0 \\ \hline
\end{tabular}
\label{tab:body_delta}
\end{table*}
\section{Versions of Articles from the arXiv.org Corpus}
About $35\%$ of all $1.1$ million papers in arXiv.org have more than one version. A new version is 
created when, for example, an author makes a change to the article and re-submits it to arXiv.org. 
The evidence of Figure \ref{fig:pub_dates} shows that the majority of the latest versions in 
arXiv.org were still uploaded prior to the publication of its final published version in a 
commercial venue. However, we were motivated to eliminate all doubt and hence decided to repeat 
our comparisons of the text contents of paper titles, abstracts, and body sections using the 
earliest versions of the articles from arXiv.org only. The underlying assumption is that those 
versions were uploaded to arXiv.org even earlier (if the authors uploaded more than one version) 
and hence are even less likely to exhibit changes due to copyediting by a commercial publisher. 
It follows, then, that if the comparisons of these earlier pre-print texts to their published 
versions show substantially greater divergences, then it is possible that more of these changes 
are the result of publisher-initiated copyediting. 

Our corpus of pre-print and final published papers matched by their DOIs and available via UCLA's 
journal subscriptions exhibits a higher ratio of papers with more than one version in arXiv.org 
than is found in the full set of articles available from arXiv.org. $58\%$ of the papers we 
compared had more than one version, $39\%$ had exactly two, and $13\%$ had exactly three versions; 
whereas only $35\%$ of all articles uploaded to arXiv.org have more than one version. We applied 
our five similarity measures (see Section \ref{subsec:txt_comp_meth}) to quantify the similarity
between the first version of all articles and their final published versions. Rather than 
repeating the histograms of Figures \ref{fig:title_histo}, \ref{fig:abstract_histo}, 
and \ref{fig:body_histo}, we show the divergences from these histograms only. Figure \ref{fig:deltas} 
displays a positive/negative barplot that represents the differences between our first comparison 
and this one. 

Figure \ref{subfig:title_delta} depicts the deltas of the title comparisons. The top bin contains 
only negative bars, meaning that our second comparison, using the earliest uploaded versions only, 
returned fewer similarity scores in that bin. The number of title comparisons in the top bin for 
the length ratio, for example, dropped by almost $1,000$ and the number in the top bin for the 
Levenshtein ratio dropped by $800$. While these numbers may at first seem dramatic, the bigger 
picture shows that the decrease is not that significant. We merely see a $10\%$ drop in the top 
bin for length ratio and Levenshtein, a $9\%$ drop for the S\o rensen and Jaccard index, and a 
drop of less than $6\%$ for the cosine similarity. The second bin shows positive bars for the 
length ratio and cosine similarity, which means that our comparison using the first uploaded 
version to arXiv.org returned more values for those measures in this bin relative to our comparison 
using the latest uploaded version. The absolute counts for the following bins decrease relative to 
our initial comparison, and it is difficult to interpret the corresponding shifts in their 
proportional values when they are plotted visually (as red diamonds). We instead detail these 
relative changes between our two sets of comparisons in Table \ref{tab:title_delta}.

The numbers for the abstract comparison are fairly similar. Figure \ref{subfig:abstract_delta} shows
a drop for all measures in the top bin and corresponding gains in the following bins. However, the
relative numbers again are not dramatic. Table \ref{tab:abstract_delta} lists all relative differences.

The results for the body comparison are interesting. As shown in Figure \ref{subfig:body_delta}, 
we observe a $18\%$ drop in length ratio and even a $43\%$ drop in Levenshtein scores for the top
bin. However, cosine scores drop by only $2\%$ in the top bin. Given that in our first body comparison
(see Section \ref{sec:body_analysis}) $80\%$ of cosine scores belonged in the top bin, the drop in this 
second body comparison is almost negligible. The detailed list of relative differences can be found in
Table \ref{tab:body_delta}.

These results confirm our initial assessment that very little difference can be found between 
pre-print articles and their final published versions. Even more so, these findings strengthen our 
argument as they show that the difference between the earliest possible pre-print version and the 
final published one seems insignificant, given the similarity measures we applied to our corpus.
\subsection{Publication Dates of Versions}
\begin{figure}[h!]
\center
\includegraphics[scale=0.17]{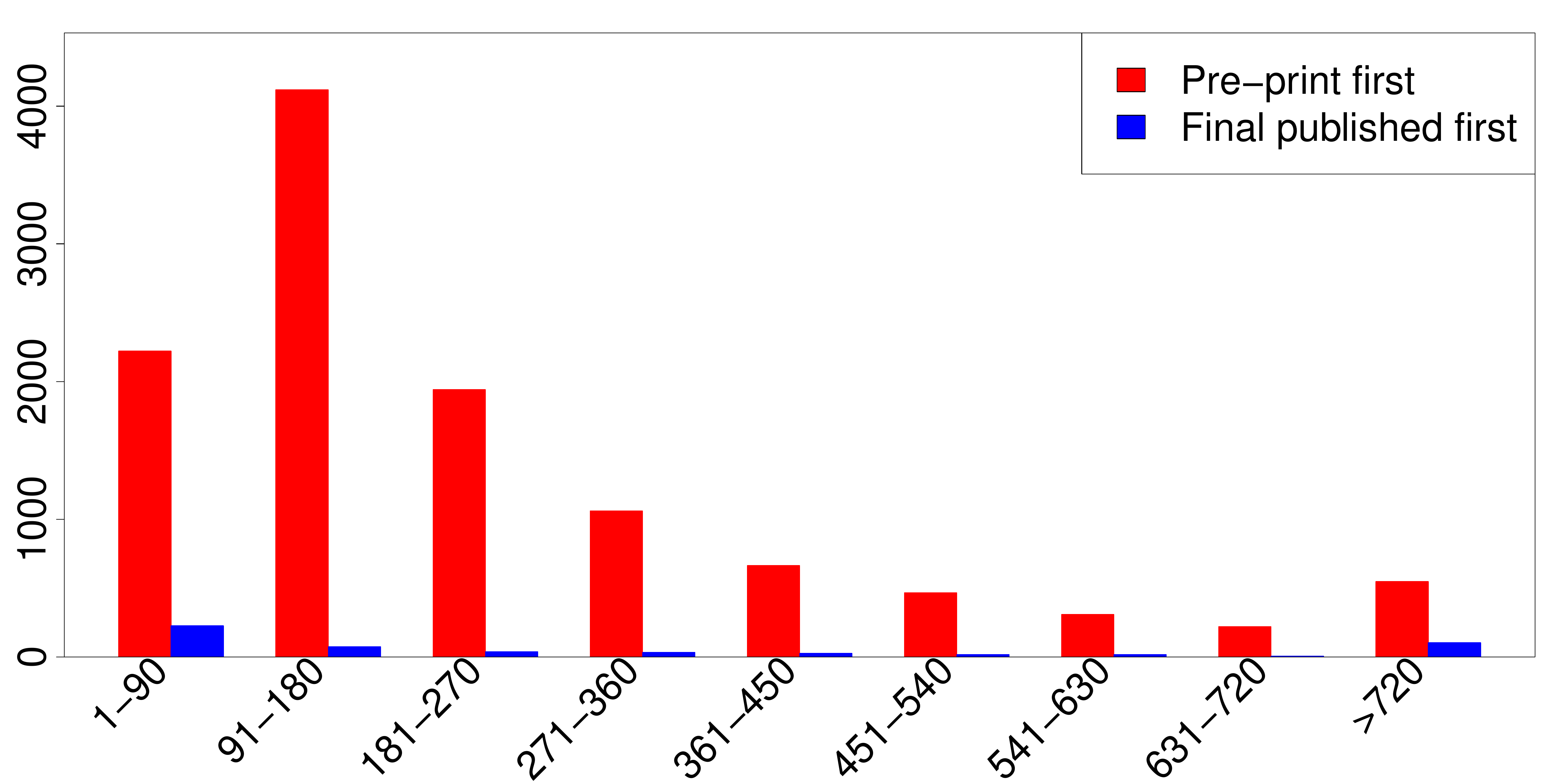}
\caption{Numbers of articles first appearing in the specified venue, given the date of the first 
pre-print upload and the commercial publication date, binned by the number of days between them}
\label{fig:pub_dates_versions}
\end{figure}

The scenarios discussed in Section \ref{subsec:pub_dates} with respect to the question of whether 
an article was uploaded to arXiv.org before it appeared in a commercial venue are valid for this 
comparison as well. 
Figure \ref{fig:pub_dates_versions} mirrors the concept of Figure \ref{fig:pub_dates} and 
shows the number of earliest pre-print versions uploaded to arXiv.org first in red
and the final published versions appearing first represented by the blue bars.
As expected, the amount of pre-print versions published first increased and now stands at $95\%$ 
as shown in Figure \ref{fig:pub_dates_versions} (compared to $90\%$ shown in 
Figure \ref{fig:pub_dates}). Our argument for the second scenario described above is therefore 
strongly supported when considering the earliest uploaded versions of pre-prints. 
\section{Discussion and Future Work}
The results outlined in this paper are from a preliminary study on the similarity of pre-print 
articles to their final published counterparts. There are many areas where this study could be 
improved and enhanced.
One limitation to this work is the focus on arXiv.org as the sole corpus of pre-print articles.
As a result, all of the articles are from a relatively limited slice of the STM domain 
--- specifically, physics, mathematics, statistics, and computer science, as shown in 
Figure \ref{fig:arxiv_categories}. Expanding this line of experiments to other domains such as the 
biological sciences, humanities, social sciences, and economics might return different results, 
as the review and editorial practices in other disciplines can vary considerably. As part of our 
future work, we are planning to conduct this experiment again with articles from the RePEc.org 
corpus (economics) and from bioRxiv.org (biology), for example.

The matching of a pre-print version of an article to its final published version was done by means 
of the article's DOI. While this is an obvious choice for a paper identifier, by only relying on 
DOIs we very likely missed out on other matching articles. For future experiments, we will include 
the paper's title and author(s) in the matching process.
Note also that we could only match articles that we have access to via the UCLA Library's serial 
subscriptions. It might be worth expanding the matching process to a collaborating organization 
with ideally complementary subscriptions to maximize access to full text articles.

One typical article section we have not analyzed as part of this research is the references section. 
Given publishers' claims of adding value to this section of a scholarly article, we are motivated 
to see whether we can detect any significant changes between pre-prints and final published versions 
there. Similarly, we have not thoroughly investigated changes in the author sections. We anticipate 
author movement, such as authors being added, being removed, and having their rank in the list of 
authors changed --- although changes in author order due to publishers' name alphabetization 
policies must be considered as well. Initial experiments in this domain have proven difficult to 
interpret, as author names are provided in varying formats and normalization is not trivial. 

Another angle of future work is to investigate the correlation between pre-prints and final 
published versions' degree of similarity and measured usage statistics such as download numbers 
and the articles' impact factor values. When arguing that the differences between pre-print articles 
and their final published versions are insignificant, factoring in usage statistics and 
``authority values'' can further inform decisions about spending on serial subscriptions.
\section{Conclusions}
This study is motivated by academic publishers' claims of the value they add to scholarly articles 
by copyediting and making further enhancements to the text. We present results from our preliminary 
study to investigate the textual similarity of scholarly pre-prints and their final published 
counterparts. We apply five different similarity measures to individual extracted sections from the 
articles' full text contents and analyze their results. We have shown that, within the boundaries 
of our corpus, there are no significant differences in aggregate between pre-prints and their 
corresponding final published versions. In addition, the vast majority of pre-prints 
($90\%$ - $95\%$) are published by the open access pre-print service first and later by a commercial 
publisher. Given the fact of flat or even shrinking library, college, and university budgets, our 
findings provide empirical indicators that should inform discussions about commercial publishers' 
value proposition in scholarly communication and have the potential to influence higher education 
and academic and research libraries' economic decisions regarding access to scholarly publications.
%
%

%
%

%
\end{document}